\documentstyle[preprint,aps]{revtex}

\begin{document}

\draft
\preprint{\parbox{5cm}{ADP-93-210/T128\\
                       PSI-PR-93-13\\
                       JUNE 1993\\[4cm]
                       \strut}}
\title{Deep Inelastic Scattering from Off-Shell Nucleons}
\author{W.Melnitchouk\footnote{Present address:
                               Institut f\"{u}r Theoretische Physik,
                               Universit\"{a}t Regensburg,
                               \mbox{D-93040} Regensburg, Germany.}}
\address{Department of Physics and Mathematical Physics,
         University of Adelaide,
         South Australia, 5005.}
\author{A.W.Schreiber}
\address{Paul Scherrer Institut,
         W\"{u}renlingen und Villigen,
         CH-5232 Villigen PSI, Switzerland.}
\author{A.W.Thomas}
\address{Department of Physics and Mathematical Physics,
         University of Adelaide,
         South Australia, 5005.}

\maketitle

\newpage
\begin{abstract}
We derive the general structure of the hadronic tensor
required to describe deep-inelastic scattering from an off-shell
nucleon within a covariant formalism.
Of the large number of possible off-shell structure functions
we find that only three contribute in the Bjorken limit.
In our approach the usual ambiguities encountered when
discussing problems related to off-shellness in deep-inelastic
scattering are not present.
The formulation therefore provides a clear
framework within which one can discuss the various approximations
and assumptions which have been used in earlier work.
As examples, we investigate scattering from the deuteron,
nuclear matter and dressed nucleons.
The results of the full calculation are compared with
those where various aspects of the off-shell structure are neglected,
as well as with those of the convolution model.
\end{abstract}

\section{Introduction}

The general structure of the hadronic tensor relevant to
deep-inelastic scattering (DIS) from an on-mass-shell nucleon
($p^2$ = $M^2$) which transforms correctly under proper Lorentz
and parity transformations, and which is gauge and time-reversal
invariant, is well known.
In the Bjorken limit the two possible structure functions collapse
to one, so that, in the case of one flavour, electromagnetic
deep-inelastic scattering may be expressed in terms of just one
quark distribution which is a function of only one variable.
(All these statements refer to spin-independent scattering,
to which we restrict ourselves throughout this paper.)

The situation is considerably more complex if one is considering,
in a covariant formulation, DIS from an off-mass-shell
($p^2$ $\ne$ $M^2$) hadronic constituent within a composite target.
This situation arises, for example, in many calculations relevant
to the EMC effect, where an off-shell nucleon contained in a nucleus
interacts with a high energy probe.
Another application of interest is the scattering from a nucleon
dressed by a meson cloud.
Indeed, because of the added complexity, many calculations ignore the
issue completely in the hope that the effects are not large.
Typically one neglects not only the possible $p^2$ dependence in the
structure functions, but also assumes no change in the structure of
the off-shell hadron tensor.
Only in this case can the structure function of the target be written
as a one-dimensional convolution between a constituent (nucleon)
distribution function within the target and a quark distribution
within the constituent \cite{JAFFE85}.

We shall consider scattering from an off-mass-shell nucleon without
making these approximations.
The purpose is to develop a theoretical framework which is exact,
thus keeping the model-dependent approximations to as late a stage
as possible.
Of course, in order to make progress we have to restrict our
consideration to the interaction with a single off-shell nucleon
(impulse approximation).
Thus processes where the lepton interacts with quarks in two or more
different nucleons (final state interactions) are excluded at this
stage, even though these may not be negligible (see Section VI B).

It is important to realise that the change in the structure of
the off-shell tensor is by no means a trivial matter.
There are several distinct differences from the on-shell tensor:
\newcounter{lis}
\begin{list}
{\Roman{lis}}{\usecounter{lis}\setlength{\rightmargin\leftmargin}}
\item \ Most obviously, the dependence on the four-momentum squared
of the nucleon is no longer trivial, as it is in the case where the
target is on-shell.
\item \ In a covariant formalism the off-shell fermion
tensor is a 4$\times$4 matrix in the external fermion legs.
This corresponds to
the fact that in a relativistic theory it is necessary to
consistently incorporate the antiparticle degrees of freedom.
Because of this matrix structure the tensor involves, at least in
principle, many more independent functions than in the on-shell case.
\item \ Because the
incoming particles are off-mass-shell the gauge invariance condition
for this tensor is not the same as in the on-shell case.
\end{list}
To show this last point, consider the truncated forward virtual
Compton amplitude, $\widehat{T}_{\mu\nu}(p,q)$, which satisfies the
well-known generalised Ward identity \cite{KAZES59}
\begin{eqnarray}
q^{\mu}\ \widehat{T}_{\mu\nu}(p,q)
& = & -e\left \{ S^{-1}(p)S(p+q)\Gamma_{\nu}(p+q,p)\ -\
\Gamma_{\nu}(p,p-q)S(p-q)S^{-1}(p)\right \},
\label{ward2}
\end{eqnarray}
where $\Gamma_{\nu}(p+q,p)$ is the $\gamma NN$ vertex function and
$S(p)$ is the fermion propagator.
For an on-shell nucleon, the full Compton amplitude is
\begin{eqnarray}
T_{\mu \nu}(p,q) = \overline {u}(p)\ \widehat {T}_{\mu \nu}\ u(p)
\label{ward1}
\end{eqnarray}
so that inserting Eq.(\ref{ward2}) into Eq.(\ref{ward1}), and using
the Dirac equation, leads to
\begin{eqnarray}
q^{\mu} T_{\mu \nu}(p,q)=0.
\label{ward1on}
\end{eqnarray}
Note that the same equation does not hold for the off-shell tensor
$\widehat{T}_{\mu\nu}$ (ie. the right hand side of Eq.(\ref{ward2})
is non-zero), even for the case where the target is a free,
pointlike fermion.

Although in calculations of nuclear structure functions the off-shell
aspects of the nucleon structure function have usually been ignored,
a few partial attempts have been made to try to account for
these effects.
Unfortunately, these calculations are not without ambiguities
\cite{JOHNSON87,HELLER90}.
Kusno and Moravcsik \cite{KUSNOMOR} used the so-called
`off-shell kinematics --- on-shell dynamics' scheme, in which the
off-shell nucleon tensor is  evaluated at the same energy
transfer $\nu$ and four-momentum transfer $q^2$ as the on-shell one,
independent of the virtuality of the nucleon.
Bodek and Ritchie \cite{BODEK81} used a similar scheme, however they
suggested that the off-shell structure functions could be identified
with the on-shell ones, evaluated for the same values of $q^{2}$
and centre of mass energy squared $s = (p+q)^{2}$, and hence a
different value of energy transfer,
$\nu \rightarrow \nu + (p^2 - M^2)/2$.
Dunne and Thomas \cite{DUNNE86}, on the other hand, used
an ansatz in which the matrix elements of the hadronic operators in
the operator product expansion were assumed to be independent
of $p^2$.
The result was a nucleon structure function that was to be evaluated
at a shifted value of $q^2 (\rightarrow \xi(p^2,q^2) q^2$,
where $\xi$ is the $q^2$-rescaling parameter).
This result was mathematically equivalent to the dynamical rescaling
model of Close, Roberts and Ross \cite{CLOSE83} and Nachtmann and
Pirner \cite{NACHTMAN84}, in which the shift in $q^2$ was attributed
to a change in confinement radius for nucleons bound inside a nucleus.

All of the above treatments use, in one form or other, the familiar
convolution formula \cite{JAFFE85}, which amounts to folding the quark
momentum distribution in the off-shell constituent
with the constituent momentum distribution in the target.
In order to derive this formula it is assumed that the
form of the off-shell nucleon tensor (i.e. the structure in its Dirac
indices) is the same as the on-shell one \cite{JUNG88,MULDERS92}.
However, as we show in Sections II and III, more than one operator
contributes in the Bjorken limit, so there is no {\em a priori}
reason for this to be a valid assumption.
The appearance of these other operator structures is closely
connected with the antiparticle degrees of freedom arising in any
relativistic treatment and constitutes an important part of the
off-shell effects.
Relativistic calculations have been attempted in the past
by Kulagin \cite{KULAGIN}, Nakano \cite{NAKANO}
and Gross and Liuti \cite{GROSS92},
however their derivations of the convolution model also relied
critically on assumptions about the off-shell tensor,
and the relativistic bound nucleon density matrix, respectively.
In fact, to our knowledge, all attempts to derive the simple
covariant convolution model have ultimately resorted to some
prescription to account for the fact that the bound nucleon has
$p^2 \not= M^2$.
Without performing a full calculation which self-consistently
accounts for the nucleon virtuality, the validity of the
various {\em ad hoc} approximations remains unclear.
In short, the naive convolution formula is not a sound starting
point for discussing off-shell effects and we make no use of it.

There exist alternative approaches to these just described
which do not suffer from off-mass-shell ambiguities.
For the nuclear EMC effect, Berger et al. \cite{BERGER84} used
light-front dynamics to calculate the nuclear structure functions.
Here all particles are on-mass-shell, the transverse momentum and the
light-cone variable $p_+ = p_0 + p_L$ are conserved at each vertex,
while $p_- = p_0 - p_L$ is not.
Alternatively, Johnson and Speth \cite{JOHNSON87} and Heller and
Thomas \cite{HELLER90} used old-fashioned
perturbation theory with
the instant form of dynamics, where particles are on-mass-shell,
three-momentum is conserved, but not necessarily energy.
Unfortunately, in both of these approaches, the off-mass-shell
ambiguities in the definition of the off-shell structure functions
are simply replaced by off-energy-shell ambiguities \cite{MA93}.
A review of some of the problems with these approaches may be found
in Refs.\cite{BICKERSTAFF89,OELFKE90}.

The advantage of the covariant method in nuclear calculations is that
Lorentz invariance is manifest.
However, for a consistent treatment within this framework one has to
include the antiparticle degrees of freedom, which has not been
done up to now.
We will set up the formalism in such a way that the
structure functions of the physical target are expressed in terms
of fully relativistic quark--nucleon and nucleon--target
vertex functions.
This will enable us to ensure gauge invariance,
the Callan-Gross relation and an unambiguous identification of the
scaling variables.
All model approximations will be contained entirely in the vertex
functions themselves, which, of course, we cannot calculate from
first principles.

This paper is organised as follows: In Section II we define the
general structure of the off-shell tensor in terms of a suitable
set of structure functions.  In Section III we explicitly calculate
the scaling properties of these functions.
As we shall see, only 3 of 14 possible functions contribute
in the Bjorken limit. In Section IV we discuss how our formalism
can be used to calculate structure functions of composite particles
and discuss the limits in which the conventional convolution model
may be obtained.
In Section V we use some simple parameterisations of the relativistic
vertex functions to calculate the nucleon valence quark
distributions.
Using these same vertex functions we then calculate in Section VI
the structure functions of composite targets containing off-shell
nucleons.

\section{General Structure of the Off-Shell Nucleon Tensor}

The process in which we are interested is depicted in Fig.\ref{1},
with the photon momentum $q$ and the off-shell nucleon momentum
$p$ marked.
Due to hermiticity, covariance, parity and time reversal invariance
the corresponding off-shell tensor $\chi_{\mu\nu}$ is a
4$\times$4 matrix depending on $p$ and $q$, and may in general be
written in terms of 14 functions:
\begin{eqnarray}
\chi_{\mu\nu}(p,q)
& = & \chi_{\mu\nu}^0(p,q)
\ + \ \not\!p\ \chi_{\mu\nu}^1(p,q)
\ + \ \not\!q\ \chi_{\mu\nu}^2(p,q)
\ + \ \gamma_{\{\mu} p_{\nu\}}\ \chi^3(p,q)
\ + \ \gamma_{\{\mu} q_{\nu\}}\ \chi^4(p,q)
\label{chiuv}
\end{eqnarray}
where the braces $\{ ... \}$ around the subscripts indicate the
symmetric $\mu\nu$ combination.
Here, $\chi_{\mu\nu}^i(p,q)$ are the most general tensors of
rank two which may be constructed out of $q$ and $p$,
\begin{eqnarray}
\chi_{\mu\nu}^i(p,q)
& = & P_{T \mu\nu}(p,q)\ \chi_T^i(p,q)
\ + \ P_{L \mu\nu}(p,q)\ \chi_L^i(p,q)                 \nonumber\\
& + & P_{Q \mu\nu}(p,q)\ \chi_Q^i(p,q)
\ + \ P_{QL \mu\nu}(p,q)\ \chi_{QL}^i(p,q),
\hspace*{1cm}  i = 0, 1, 2.                         \label{eq:tensor}
\end{eqnarray}
The functions $\chi^i$ on the right hand side of Eq.(\ref{eq:tensor}),
as well as $\chi^3$ and $\chi^4$ in Eq.(\ref{chiuv}),
are real scalar functions of $q$ and $p$.
The tensors $P^{\mu\nu}$ are defined by
\begin{eqnarray}
P^{\mu\nu}_T(p,q)
&=& \widetilde{g}^{\mu\nu}
 + { \widetilde{p}^{\mu} \widetilde{p}^{\nu} \over \widetilde{p}^{2} },
\hspace*{1.5cm}
P^{\mu\nu}_L(p,q)\
=\ { \widetilde{p}^{\mu} \widetilde{p}^{\nu} \over \widetilde{p}^{2} },
\nonumber\\
P^{\mu\nu}_Q(p,q)
&=& \frac{ q^{\mu} q^{\nu} }{ q^{2} },
\hspace*{1.5cm}
P^{\mu\nu}_{QL}(p,q)\
=\ { 1 \over \sqrt{-q^{2} \widetilde{p}^{2}} }
 \left( \widetilde{p}^{\mu} q^{\mu} + \widetilde{p}^{\nu} q^{\mu}
 \right),
\label{proj}
\end{eqnarray}
where $\widetilde{g}_{\mu\nu} = -g_{\mu\nu} + q_{\mu} q_{\nu} / q^2$
and $\widetilde{a}_{\mu} = a_{\mu} - q_{\mu}\ a \cdot q/q^2$,
with $a_{\mu}$ being any four-vector.

The above decomposition of the off-shell tensor is of course not
unique.
It is written in this convenient form because the tensors
$P^{\mu\nu}$ turn out to be projection operators
\cite{MULDERS92} and satisfy
\begin{eqnarray}
P^{\mu\nu}_T(p,q)\ P_{T \mu\nu}(p,q) &=& 2,  \hspace*{1.5cm}
P^{\mu\nu}_L(p,q)\ P_{L \mu\nu}(p,q)\ =\ 1,  \nonumber\\
P^{\mu\nu}_Q(p,q)\ P_{Q \mu\nu}(p,q) &=& 1,  \hspace*{1.5cm}
P^{\mu\nu}_{QL}(p,q)\ P_{QL \mu\nu}(p,q)\ =\ -2,
\end{eqnarray}
with all other combinations vanishing.
It is important to note that in the Bjorken limit these relations are
also true for projectors involving different momenta. That is,
the projectors are still orthogonal in this limit and
\begin{eqnarray}
P^{\mu\nu}_T(p_1,q)\ P_{T \mu\nu}(p_2,q) &=&  2\;\;\;{\rm etc.}
\end{eqnarray}

In general, Fig.\ref{1} is a subdiagram of Fig.\ref{2},
where ${\cal P}$ is the on-shell
momentum of the composite target (labelled $A$).
As will be discussed more fully in Sections IV and V, the hadron
tensor for the
complete process, $W^A_{\mu\nu}({\cal P},q)$, involves an integral
over the nucleon momentum $p$ of the tensor $\chi_{\mu\nu}(p,q)$,
traced with another 4$\times$4 matrix originating in the soft
target--constituent part of the diagram.
Hence no experiment measures the off-shell tensor by itself, so it is
not possible to measure all the functions $\chi^i$ separately.
Only combinations thereof give rise to observable experimental
quantities.
Using the above projectors, we can determine which combinations of
off-shell structure functions contribute to the physical ones.
In particular, the operators $P_T^{\mu\nu}({\cal P},q),\
P_L^{\mu\nu}({\cal P},q),\
P_Q^{\mu\nu}({\cal P},q)$ and
$P_{QL}^{\mu\nu}({\cal P},q)$ project from the composite target
tensor $W^A_{\mu\nu}({\cal P},q)$ the transverse,
longitudinal and the two possible gauge non-invariant contributions,
respectively, in terms of the scalar functions $\chi^i(p,q)$.

Not all of the functions $\chi^i$ will in fact be independent,
as the gauge invariance of the theory requires that the latter two
contributions vanish.
Furthermore, the longitudinal function must also be zero in the
Bjorken limit
(${\cal P} \cdot q,\ Q^2 \equiv -q^2 \rightarrow \infty,\
x = Q^2/2{\cal P}\cdot q$ fixed),
if the Callan-Gross relation is to be satisfied.
That this is indeed the case is shown explicitly in the Appendix.
For the remaining physical (transverse) contribution we obtain for
the coefficients of the $\chi^i$'s:
\begin{eqnarray}
\frac{1}{2} P^{\mu\nu}_T({\cal P},q)\ \chi_{\mu\nu}(p,q)
&=& \chi^0_T(p,q)\
 +\ \not\!p\ \chi^1_T(p,q)\
 +\ \not\!q\ \chi^2_T(p,q)\                               \nonumber\\
&+&
{q^2 \over 2(p \cdot q)^2}\ \left( p - y {\cal P} \right)^2
\left[ \chi^0_L(p,q)\
 +\ \not\!p\ \chi^1_L(p,q)\ +\ \not\!q\ \chi^2_L(p,q)\        \right]
                                                          \nonumber\\
&+&
\left[ - \not\!p + y \not\!\!{\cal P}
          + {1\over {\cal P} \cdot q} \left( {\cal P} \cdot p
                                      - y {\cal P}^2 \right) \not\!q
   \right] \chi^3(p,q)                                 \label{prwh}
\end{eqnarray}
where $y = p \cdot q / {\cal P} \cdot q  = p_+ / {\cal P}_+$ is
the constituent's light-cone
momentum fraction.
In the next Section we derive the scaling behaviour of the
functions $\chi^i$ using the parton model, by separating the hard,
$q^2$-dependent part of the truncated amplitude
$\chi_{\mu\nu}(p,q)$ from the soft, non-perturbative component.
We will see that Eq.(\ref{prwh}) simplifies considerably,
as many terms do not contribute in the Bjorken limit.

A special case of the above formalism is DIS from an on-shell
nucleon, described by the tensor which we denote
by $W^N_{\mu\nu}(p,q)$.
In this case the contribution to the nucleon tensor is given by
Eq.(\ref{prwh}) traced with $ (\not\!p + M)/2$, where
${\cal P}$=$p$ and no integration over $p$ is performed:
\begin{eqnarray}
M\ W^N_{\mu\nu}(p,q;p^2=M^2)
&=& {1 \over 2} {\rm Tr} \left[ (\not\!p + M)\
                         \widetilde{\chi}_{\mu\nu}(p,q) \right].
\label{on}
\end{eqnarray}
This gives the transverse unpolarised on-shell structure functions
in terms of the on-shell limits of the functions $\chi^i$
($\widetilde{\chi}^i(p,q) \equiv \chi^i(p,q;p^2=M^2)$):
\begin{eqnarray}
{M \over 2} W^N_T(p,q)
&=& M\ \widetilde{\chi}^0_T(p,q)\ +\ M^2\ \widetilde{\chi}^1_T(p,q)\
                              +\ p \cdot q\
    \widetilde{\chi}^2_T(p,q).
\label{WTreal}
\end{eqnarray}
Similar expressions can also be found for the other functions
(i.e. longitudinal and non gauge-invariant), but again these vanish
in the Bjorken limit.

\section{Scaling Behaviour of the Off-Shell Functions $\chi$}

In this Section we calculate the leading twist contribution to the
off-shell structure functions within the covariant quark-parton model.
The formal result of the operator product expansion enables us to
separate the imaginary part of the forward scattering amplitude,
depicted in Fig.\ref{3},
into its hard (calculable perturbatively) and soft (non-perturbative)
components, denoted by $r_{\mu\nu}$ and $H(k,p)$, respectively.
A method similar to this was also discussed in Ref.\cite{ELLIS83}.
This then enables the off-shell tensor $\chi_{\mu\nu}$ to be
written as:
\begin{eqnarray}
\left[ \chi_{\mu\nu}(p,q)\right]_{ab}
= {\cal I} \cdot \left[ r_{\mu\nu}(k,q) \right]_{cd}
  \left[ H(k,p) \right]_{dcab}                   \label{eq:initial}
\end{eqnarray}
where ${\cal I}$ is the integral operator
\begin{eqnarray}
{\cal I}
 = \int {d^4k \over (k^2-m^2)^2} \delta([q+k]^2-m^2)    \label{Iop}
\end{eqnarray}
and the Dirac matrix structure has been made explicit.
(The complete forward scattering amplitude in addition contains the
crossed photon diagram, which we do not explicitly take into account.
All the formal results of Sections I to IV remain valid upon
inclusion of this diagram.
Numerically, it can make a small contribution in the small-$x$
region, however in the subsequent model calculation in which we
consider only two-quark intermediate states there will be no
contribution.)
In Eq.(\ref{Iop}) $k$ is the parton's four-momentum and $m$ its
(current) mass.
In the following we will drop quark mass terms as the difference
between the $m=0$ results and those for $m \sim$ few MeV is
negligible.
(We shall return to the question of quark masses in Section V.)
The vector nature of the quark--photon coupling then determines
the structure of the tensor $r_{\mu\nu}$ to be:
\begin{eqnarray}
r_{\mu\nu}(k,q)
= k^2 \left( q_{\alpha}g_{\mu\nu} -
             \left( k_{\{\mu}+q_{\{\mu} \right) g_{\nu\}\alpha}
      \right) \gamma^{\alpha}
\ +\ \not\!k \left( q^2 g_{\mu\nu} + 4 k_{\mu} k_{\nu}
                 + 2 k_{\{\mu}q_{\nu\}} \right).
\end{eqnarray}
Following this, the trace over the indices $c, d$ in
Eq.(\ref{eq:initial}) may be performed and the results
written as:
\begin{eqnarray}
{\rm Tr} \left[ r_{\mu\nu} [H]_{ab} \right]
= \left[ k^2 \left( q_{\alpha} g_{\mu\nu}
       - \left( k_{\{\mu} + q_{\{\mu} \right) g_{\nu\}\alpha}
         \right)
       + k_{\alpha} \left( q^2 g_{\mu\nu} + 4 k_{\mu} k_{\nu}
                         + 2 k_{\{\mu} q_{\nu\}}
                    \right)
  \right]
[G^{\alpha}(p,k)]_{ab}.
\end{eqnarray}
As $G^{\alpha}$ is a 4$\times$4 matrix which transforms like a vector
and must be even under parity transformations, its most general
form is
\begin{eqnarray}
G^{\alpha}
&=& I \left( p^{\alpha} f_1 + k^{\alpha} f_2 \right)\
+\ {\not\!k} \left( p^{\alpha} f_3 + k^{\alpha} f_4 \right)\
+\ {\not\!p} \left( p^{\alpha} f_5 + k^{\alpha} f_6 \right)\
+\ \gamma^{\alpha} f_7,
\label{Galpha}
\end{eqnarray}
where the functions $f_i$ are scalar functions of $p$ and $k$.

The integrals over $k$ can be done in a standard way.
For example, for an integrand containing one free $k^{\alpha}$,
contracting with $p_{\alpha}$ and $q_{\alpha}$ enables us to make
the replacement
\begin{eqnarray}
{\cal I}\cdot k^{\alpha} \rightarrow {\cal I}\cdot\left\{
 \rho_{1} \ p^{\alpha}+\rho_2 \ q^{\alpha} \right\}.
\end{eqnarray}
Similarly for $k^{\alpha} k^{\beta}$ terms,
\begin{eqnarray}
{\cal I}\cdot k^{\alpha} k^{\beta} \rightarrow {\cal I}\cdot\left\{
\rho_3 \ P^{\alpha\beta}_T(p,q)+
\tilde{p}^2 \rho_1^2\ P^{\alpha\beta}_L(p,q)+
{(k \cdot q )^2 \over q^2} P^{\alpha\beta}_Q(p,q)-
{\tilde{k} \cdot \tilde{p} \ k \cdot q \over
 \sqrt{-q^2 \ \tilde{p}^2}} P^{\alpha\beta}_{QL}(p,q)\right\}
\end{eqnarray}
and for $k^{\alpha} k^{\beta} \not\!k$ terms,
\begin{eqnarray}
{\cal I} \cdot k^{\alpha} k^{\beta} \not\!k
&\rightarrow&
 {\cal I} \cdot \left\{
- \rho_3 \left( \rho_1 p^{\{\alpha}
+ \rho_2 q^{\{\alpha} \right) \gamma^{\beta\}}
+ \rho_3 \left( \rho_1 \not\!p + \rho_2 \not\!q \right)\
  P^{\alpha\beta}_T(p,q)
\right.  \nonumber\\
&+&
\rho_1 \left( \left[ 2 \rho_3+\tilde{p}^2 \rho_1^2 \right] \left[
\not\!p-{p \cdot q \over q^2} \not\!q \right] + {k \cdot q \over q^2}
  \tilde{p}^2 \rho_1 \not\!q\right)\
P^{\alpha\beta}_L(p,q)\\
&+&
{(k \cdot q )\over q^2} \left( k \cdot q \rho_1 \not\!p +
\left[k \cdot q \rho_2+ 2 \rho_3\right] \not\!q \right)\
P^{\alpha\beta}_Q(p,q)
\nonumber\\
&-& \left.
{1 \over \sqrt{-q^2 \ \tilde{p}^2}} \left(
\left[k \cdot q \rho_3 + \tilde{p}^2 \rho_1^2 \right] \not\!p +
\left[\tilde{p}^2 \rho_1 (\rho_3+k \cdot q \rho_2)
     - {p \cdot q\  k \cdot q \over q^2} \rho_3\right]
\not\!q \right) P^{\alpha\beta}_{QL}(p,q) \right \},
\nonumber
\end{eqnarray}
where
\begin{eqnarray}
\rho_1&=&{\tilde{k} \cdot \tilde{p}  \over \tilde{p}^2} \nonumber\\
\rho_2&=&{k \cdot q \over q^2}
          - {p \cdot q \ \tilde{k} \cdot \tilde{p}
                \over q^2\ \tilde{p}^2} \\
\rho_3&=&{1 \over 2}
         \left(-\tilde{k}^2 + \tilde{p}^2 \rho_1^2\right) \nonumber.
\end{eqnarray}

The $\chi^i$ are then completely defined in terms of the functions
$f_1$--$f_7$, and as all the dependence on the photon momentum $q$
is now explicit, their scaling behaviour may be derived in a
straightforward manner.
We find that $\chi^0_T$ and $\chi^1_T$ are of order 1, while all
other $\chi^i$ are of order $1/\nu$.
Hence we find that deep-inelastic scattering from an off-shell
nucleon may be expressed in terms of just three functions,
\begin{eqnarray}
\frac{1}{2} P^{\mu\nu}_T({\cal P},q)\ \chi_{\mu\nu}(p,q)
&=& \chi^0_T(p,q)\ +\ \not\!p\ \chi^1_T(p,q)\
 +\ \not\!q\ \chi^2_T(p,q).                          \label{prwhs}
\end{eqnarray}

The complete expressions for the functions $\chi_T^i$ are:
\begin{mathletters}
\label{chiT}
\begin{eqnarray}
\chi^0_T(p,q) &=&
{\cal I} \cdot \left\{ -\left(k^2 p \cdot q + q^2 k \cdot p\right)
f_1(k,p)- {q^2 k^2 \over 2} f_2(k,p) \right\}                    \\
\chi^1_T(p,q) &=& {\cal I} \cdot \left\{ - \left(k^2 p \cdot q
                                         + q^2 k \cdot p\right)
 \left( f_5(k,p)
      - {q^2 \over 2 p \cdot q} f_3(k,p) \right) \right. \nonumber\\
& & \left. \hspace*{1cm}
- {q^2 k^2 \over 2} \left(f_6(k,p)
                   -{q^2 \over 2 p \cdot q} f_4(k,p) \right) +
{q^4 \over 2 p \cdot q} f_7(k,p)\right\}                        \\
\chi^2_T(p,q) &=& {\cal I} \cdot \left\{
-\left( {p^2 q^2 \over 2p \cdot q} + k \cdot p \right)
 \left( \left[ k^2+{k \cdot p \ q^2 \over p \cdot q} \right] f_3(k,p)
             + {q^2 \over 2 p \cdot q} \left[ k^2\ f_4(k,p)
                                              + 2\ f_7(k,p)
        \right]
 \right)
 \right.                                                \nonumber\\
& & \hspace*{3cm} \left. -\ k^2\ f_7(k,p)
\right\}.
\end{eqnarray}
\end{mathletters}
For the other $\chi$ it can be easily demonstrated
(see the Appendix for details) that for each of the arbitrary
functions $f_i$, there are cancellations at leading order in $\nu$
in the expressions for
$P_L^{\mu\nu}({\cal P},q) \chi_{\mu\nu}(p,q)$,\ \
$P_Q^{\mu\nu}({\cal P},q) \chi_{\mu\nu}(p,q)$\ and\
$P_{QL}^{\mu\nu}({\cal P},q) \chi_{\mu\nu}(p,q)$\
in the Bjorken limit.
Hence the Callan-Gross relation, as well as gauge invariance
($q^{\mu} W^A_{\mu\nu}=0$), are assured, independent of the
nature of the target--constituent part of the diagram.
This result is completely general, so that model dependent
approximations for the vertex functions do not affect these
results.

\section{Convolution Model}

Before we move on to making model dependent assumptions for the
vertex functions, we need to write down the on-shell tensor
$W^A_{\mu \nu}({\cal P},q)$ for the target $A$ in terms of the
off-shell tensor $\chi_{\mu \nu}(p,q)$.
The full tensor for the composite target is given by
\begin{eqnarray}
M_T\ W^A_{\mu\nu}({\cal P},q)
&=&
\int { d^4p \over (2\pi)^4 }
{ 2\pi \delta \left( [{\cal P}-p]^2 - M_R^2 \right)
 \over (p^2 - M^2)^2 }\
{\rm Tr} \left[ \left( I A_0(p,{\cal P})
                + \gamma_{\alpha} A_1^{\alpha}(p,{\cal P})
                \right)\
\chi_{\mu\nu}(p,q) \right]                             \label{ccc}
\end{eqnarray}
where $A_0$ and $A_1$ are functions describing the
target---constituent part of the complete diagram in Fig.\ref{2},
and $M_T$ and $M_R$ are the masses of the target and target recoil
systems, respectively.
Note that Eq.(\ref{ccc}) applies to a target recoil state of
definite mass, $M_R$.
In general we could also have a sum over all excited recoil
states, or equivalently an integration over the masses $M_R$
weighted by some target recoil spectral function.
The transverse structure function of the target is obtained
from Eq.(\ref{ccc}) by using the transverse projection operator
defined in Eq.(\ref{proj}):
\begin{eqnarray}
& & M_T\ W^A_T({\cal P},q)\
=\ {M_T \over 2} P_T^{\mu\nu}({\cal P},q)\ W^A_{\mu\nu}({\cal P},q)
                                                         \nonumber\\
&=&
{1\over 4\pi^2} \int { dy\ dp^2 \over (p^2 - M^2)^2 }
 \left\{ A_0(p,{\cal P})\ \chi^0_T(p,q)\
      +\ p \cdot A_1(p,{\cal P})\ \chi^1_T(p,q)\
      +\ q \cdot A_1(p,{\cal P})\ \chi^2_T(p,q)
 \right\}
\label{targetstruc}
\end{eqnarray}
where $p^2 = p_+ p_- - {\bf p}_T^2$,\ \
$p_+ = y {\cal P}_+ (= y M_T$ in the target rest frame),
and we have used the $\delta$-function to fix
$p_- = M_T\ +\ (M_R^2 + {\bf p}_T^2)/(p_+ - M_T)$.

The convolution model may only be derived from
Eq.(\ref{targetstruc}) if we make some additional assumptions.
First of all, we need to assume that the target structure function
can be written in factorised form,
in terms of the nucleon structure function,
$W^N_T$, and some nucleon distribution function, $\varphi$:
\begin{eqnarray}
W^A_T(x,Q^2)
&=& \int dy\ \int dp^2\ \ \varphi(A_0,A_1)\ \
    W^N_T(x/y,Q^2,p^2).
\label{con1}
\end{eqnarray}
Furthermore, to obtain the usual one-dimensional convolution formula
\cite{JAFFE85,MULDERS92} we must assume that $W^N_T$
is independent of $p^2$:
\begin{eqnarray}
W^A_T(x,Q^2)
&=& \int dy\ \widetilde{\varphi}(y)\ \ W^N_T(x/y,Q^2),
\label{con2}
\end{eqnarray}
where now the integral over $p^2$ has been absorbed into
the definition of $\widetilde{\varphi}$.

There are several ways in which the first assumption might be valid:

CASE (a):\ If all but one of the functions $\chi_T^i\ (i=0-2)$
are zero in the Bjorken limit.
Most authors (see for example Refs. \cite{JAFFE85,JUNG88,MULDERS92})
adopt this choice, as this is the case for a pointlike fermion
(where only $\chi_T^2$ contributes).
However, as was shown in Section III, all three functions
$\chi_T^i$ in principle
contribute in the Bjorken limit, so that one would require that
some of the functions $f$ in Eqs.(\ref{chiT}) vanish or cancel.
We know of no reason why this should be the case --- indeed even
the extremely simple quark--nucleon vertex
functions which we consider in the next Section give rise
to more than one non-vanishing $\chi_T^i$.

CASE (b):\ If more than one of the $\chi_T^i$ is non-zero, but the
non-zero ones are proportional to each other.
For example, $f_1 = M f_5$ and all other $f$'s equal to zero would
imply $\chi_T^0 = M \chi_T^1$, and so Eq.(\ref{con1}) is obtained.
Again, in general there doesn't seem to be any reason to expect
this behaviour.

CASE (c):\ If the non-zero nucleon--target functions $A_{0,1}$
multiplying the functions $\chi_T^i$ are proportional to each other.
An example of this would be if
$A_0\ =\ p\cdot A_1 / M\ =\ q\cdot A_1\ M/p\cdot q$,
which would then give Eq.(\ref{con1}).
In general this will not be true unless the $p^2 = M^2$ limit is
taken inside the functions $A_{0,1}$.

In short, none of the above conditions are generally satisfied
in a self-consistent, fully covariant (relativistic) calculation.
Consequently the convolution model interpretation, Eq.(\ref{con2}),
of the nuclear structure function in terms of bound nucleon structure
functions is inconsistent within this formalism.
This difficulty is intrinsically related to the presence of
antinucleon degrees of freedom, which are not accounted for
in the traditional convolution model.
Furthermore, in the absence of the convolution model, the common
practice of extracting nucleon structure functions from
nuclear DIS data is rather ambiguous.
Indeed, the very concept of a structure function of a nucleon bound
within a nucleus looses its utility.
One is {\em forced} to consider quark and nuclear degrees of
freedom side by side in the calculation of nuclear structure
functions.

Using Eq.(\ref{targetstruc}) directly we may compare, for some simple
vertex functions, the exact result with those obtained by making the
convolution model approximation, Eq.(\ref{con2}).
This we will do in the next Section.
As a final comment, it should be noted that, within the physical
assumptions made by the use of the model in the first place
(i.e. no final state interactions), the functions $\chi_T^i$ are
independent of the physical target, and depend only on the
constituent nucleon.
By selecting various targets (i.e. by varying $A_{0,1}$) the
relative contributions from the functions $\chi^i_T$ could in
principle be probed, provided, of course, we know the
nucleon--target functions sufficiently well.
Conversely, once the $\chi_T^i$ have been determined for one process,
they may be used for all other processes.

\section{Calculation of the Nucleon Structure Function}

To calculate the transverse structure function of the complete
target requires two sets of functions describing the soft,
non-perturbative physics, namely the quark--nucleon functions
$f_1$--$f_7$, and the nucleon--target functions $A_0, A_1$.
Here we concentrate on the former set.

We observe that because both the constituent nucleon and
struck quark inside the nucleon have spin 1/2, the intermediate
spectator state will have either spin 0 or 1.
In order to make an overall Lorentz scalar, we therefore need
only consider quark--nucleon vertices that transform as a
scalar or vector under Lorentz transformations.
It is straightforward to identify the form of the vertices that are
allowed by Lorentz, parity and time-reversal invariance, however the
specific momentum dependence has to be determined within a model.
There will be 15 independent scalar ($\Phi^S_{1-4}(k,p)$)
and vector ($\Phi^V_{1-11}(k,p)$) vertex functions appearing in
the general expression
\begin{eqnarray}
{\cal V}^S & = &
   I\ \Phi^S_1\
+\ \not\!p\ \Phi^S_2\
+\ \not\!k\ \Phi^S_3\
+\ i \sigma_{\alpha\beta} p^{\alpha} k^{\beta}\ \Phi^S_4
\end{eqnarray}
for a scalar vertex, and
\begin{eqnarray}
{\cal V}^V_{\alpha} & = &
  \gamma_{\alpha}\ \Phi^V_1\
+\ p_{\alpha}\ I\ \Phi^V_2\
+\ k_{\alpha}\ I\ \Phi^V_3\
+\ i \sigma_{\alpha\beta} p^{\beta}\ \Phi^V_4\
+\ i \sigma_{\alpha\beta} k^{\beta}\ \Phi^V_5\      \nonumber\\
& + &
   p_{\alpha} \not\!p\ \Phi^V_6\
+\ p_{\alpha} \not\!k\ \Phi^V_7\
+\ k_{\alpha} \not\!p\ \Phi^V_8\
+\ k_{\alpha} \not\!k\ \Phi^V_9\                   \\
& + &
   i \sigma_{\beta\delta} p^{\beta} k^{\delta}\ p_{\alpha}\
   \Phi^V_{10}\
+\ i \sigma_{\beta\delta} p^{\beta} k^{\delta}\ k_{\alpha}\
   \Phi^V_{11}
\nonumber
\end{eqnarray}
for a vector vertex.

The functions $f_1$--$f_7$ in Eq.(\ref{Galpha}) can be
uniquely determined from these vertex functions.
To see this, let us firstly consider the scalar vertex.
The general target--constituent function from Section III,
$(H(k,p))_{d c a b}$, will be proportional to
$({\cal V}^{S \dagger})_{a c} ({\cal V}^S)_{d b}$.
Using the Fierz theorem the Dirac indices can be rearranged into a
form that enables the connection with the functions $f_i$ to be
explicit:
\begin{eqnarray}
f_1 &=& 2\ \left( \Phi_1^S\ \Phi_2^S\
                -\ k \cdot (p\ \Phi_2^S\ +\ k\ \Phi_3^S)\ \Phi_4^S
           \right) \delta                               \nonumber\\
f_2 &=& 2\ \left( \Phi_1^S\ \Phi_3^S\
                +\ p \cdot (p\ \Phi_2^S\ +\ k\ \Phi_3^S)\ \Phi_4^S
           \right) \delta                               \nonumber\\
f_3 &=& 2\ \left( \Phi_2^S\ \Phi_3^S\
                +\ p \cdot k\ (\Phi_4^S)^2
                -\ \Phi_1^S\ \Phi_4^S
           \right)\ \delta                              \nonumber\\
f_4 &=& 2\ \left( (\Phi_3^S)^2\ -\ p^2\ (\Phi_4^S)^2
           \right)\ \delta                        \label{fscalar}\\
f_5 &=& 2\ \left( (\Phi_2^S)^2\ -\ k^2\ (\Phi_4^S)^2
           \right)\ \delta                              \nonumber\\
f_6 &=& 2\ \left( \Phi_2^S\ \Phi_3^S\
                +\ p \cdot k\ (\Phi_4^S)^2
                +\ \Phi_1^S\ \Phi_4^S
           \right)\ \delta                              \nonumber\\
f_7 &=& \left( (\Phi_1^S)^2\ -\ (p\ \Phi_2^S\ +\ k\ \Phi_3^S)^2\
               +\ (p^2 k^2 - (p \cdot k)^2)\ (\Phi_4^S)^2
        \right)\ \delta                                 \nonumber
\end{eqnarray}
where $\delta \equiv \delta \left( [p-k]^2-m_S^2 \right)$ and
$m_S$ is the mass of the
scalar spectator system.

Calculating the functions $\Phi^S_{1-4}$ from first principles
amounts to solving the relativistic, many-body bound state problem.
As this is presently not possible one could resort to
models such as the MIT bag model.
It is not our aim to do this in this paper.
Rather, we shall choose a single scalar vertex,
say ${\cal V}^S = I\ \Phi_1^S$,
and use phenomenological input to constrain its
functional form. From Eq.(\ref{fscalar}) we find that the
$I\ \Phi_1^S$ vertex contributes only to $f_7$:
\begin{eqnarray}
f_7 = (\Phi_1^S)^2\ \delta \left( [p-k]^2-m_S^2 \right)
\hspace{3cm} [\rm scalar\ vertex].
\label{sca}
\end{eqnarray}
Similarly we choose for the vector vertex a single form,
${\cal V}^V_{\alpha} = \gamma_{\alpha} \Phi_1^V$,
and find that this makes the following contributions:
\begin{eqnarray}
f_3 &=& -f_4\ =\ -f_5\ =\ f_6\
     =\ -{2\ f_7 \over m_V^2}\
     =\ -{2\ (\Phi_1^V)^2\ \delta \left( [p-k]^2-m_V^2 \right)
                        \over m_V^2}\ \ \ \ [\rm vector\ vertex]
\label{vec}
\end{eqnarray}
where $m_V$ is the mass of the vector spectator state.
In writing Eq.(\ref{vec}) we have assumed that the intermediate
vector state has a Lorentz structure
$-g_{\alpha\beta}\ +\ (p_{\alpha} - k_{\alpha})
                      (p_{\beta} - k_{\beta})/m_V^2$.
For the sake of simplicity we further assume that only valence
quarks are present, so that the scalar or vector spectator may be
identified with a diquark.
In a more refined calculation one could, for example, integrate over
diquark masses using some diquark spectral function.

{}From Eqs.(\ref{sca}) and (\ref{vec}) we see that even the simplest
vertex functions lead to a large number of non-zero functions $f_i$.
This in turn implies that there are scaling contributions to both of
the functions $\chi_T^1$ and $\chi_T^2$ in Eqs.(\ref{chiT}), thereby
failing to satisfy scenarios (a) and (b) in Section IV for the
derivation of the convolution model.
For more complicated quark--nucleon vertices, even more of the
$f$'s will be non-zero.

The $k^2$ dependence of the functions $\Phi_1^{S,V}$ can be most
easily modelled by considering the on-shell nucleon structure
function.
(We shall approximate the quark's off-shell dependence to be the
same in on- and off-shell nucleons --- see Section VI A for a
further discussion on this point.)
The large-$x$ limit is known to be dominated by
valence $u$ quarks, which implies
that the scalar vertex dominates at large $x$ \cite{CT88}.
Now note that, as the spectator state is on-mass-shell, the quark
four-momentum will behave as
$k^2 \sim (-m_S^2 - {\bf k}_T^2) / (1-x)$
at large $x$.
In order to obtain the correct large-$x$ behaviour of the
structure function,
namely $W^N_T \sim (1-x)^3$, the $k^2$ dependence in the scalar
vertex function must be $1/k^2$, after we also take into account
the two quark propagators, as well as the factor $(1-x)$
arising from the delta function
$\delta \left( [p-k]^2-m_R^2 \right)$ for the on-shell diquark state
of mass $m_R\ (= m_S$ or $m_V)$ \cite{MEYER91}.

We fix the large-$k^2$ behaviour of the vector vertex function in a
similar way, this time by requiring that we obtain the correct
valence $d_V/u_V$ ratio at large $x$, namely $\sim (1-x)$.
This means that the vertex function for DIS from valence $d_V$ quarks
has to go like $(1-x)^4$ for large $x$, i.e. like $(k^2)^{-5/2}$
(there is an additional $(1-x)^{-2}$ factor arising from
the trace for the vector diquark).

It may now seem reasonable to choose a simple monopole form for the
scalar vertex function,
as was done, for example, in Refs.\cite{MEYER91,MULDERS92},
and a corresponding one for the vector vertex.
We do not do this, however, for the following reason.
The quark propagator, $(k^2-m^2)^{-2}$, in Eq.(\ref{Iop}) contains
a pole.
Because the kinematic maximum for $k^2$ is $(M-m_R)^2$, this pole is
in the physical region of $k^2$ when $m_R + m < M$.
The origin of this pole is clear --- the model, so far, is not
confining and the proton may dissociate into its quark and
diquark constituents.
One solution would be to make the sum of the quark and
diquark masses so large that this cannot occur.
However, we do not believe that this is desirable --- confinement
occurs not because the quark mass is large (it is only a few MeV),
but in a dynamical way associated with the nature of the colour
interaction.
The only way that the information about colour confinement can
enter in this model is through the relativistic quark--nucleon
vertex function.
A convenient way to ensure that the contribution from a deconfined
quark is excluded is to choose a numerator in $\Phi_1^{S,V}$ so that
the integrand in the structure function remains finite at the
on-shell point, $k^2 = m^2$.

For the masses of the scalar and vector diquark, $m_S$ and $m_V$,
the only information available to us is that from low energy models,
such as the bag model or the non-relativistic quark model.
There, at a scale ($Q^2$) of order a few hundred MeV$^2$,
the diquark masses are expected to be somewhere within the range
of 600 to 1100 MeV \cite{CT88,SST91}.
Furthermore, from the nucleon---$\Delta(1232)$ mass splitting we
also anticipate that $m_V$ would be some 200 MeV larger than $m_S$.

The $p^2$ dependence of the vertex functions is of course more
difficult to obtain, since for this purpose data on nuclear structure
functions must be used.
In this case the $p^2$ dependence will not be restricted to the
quark--nucleon vertex function alone, but will also be present
in the nucleon--nucleus vertices, which introduces an inherent
uncertainty in the determination of the former.
Nevertheless, the functions $\Phi_1^{S,V}$ do not depend
on the nuclear target --- that information is contained entirely
in the functions $A_{0,1}$.
Since for the deuteron the $p^2$ dependence of the relativistic $DNN$
vertex can be related to known deuteron wavefunctions
\cite{BUCK79,LOCH91,GARI92}, we may use deuteron DIS data
to constrain this universal $p^2$ dependence of the quark--nucleon
vertex functions.

In order to obtain the valence quark distribution for the deuteron,
we will use data obtained from muon
scattering for $x > 0.3$, where valence quarks are known to dominate.
Because of isospin symmetry ($u^D = d^D$) only a single experimental
quantity for the deuteron
(compared with two--- $u+d$ and $d/u$ ---for the nucleon)
is meaningful,
namely $F_{2D} =  x (4 u^D + d^D) / 9 = 5 x (u + d) / 9$,
where $u^D$, $d^D$ and $u$, $d$ are the up and down quark
distributions in the deuteron and bound proton, respectively.
Hence we cannot differentiate between the $p^2$ dependence in
$\Phi_1^S$ and that in $\Phi_1^V$.
We therefore choose a simple monopole form
and use the same cut-off mass, $\Lambda_p$, in both functions.
A detailed comparison between the model and data for $x \alt 0.3$
would require separation of the valence and sea components of
$F_{2D}$.
Although in principle this could be done by analysing
the $\nu-D$ and $\bar{\nu}-D$ DIS data,
in practice those data suffer from poor statistics.
Furthermore,
typically only the extracted quark distributions in the nucleon are
presented \cite{BEBC85}, and these depend on the theoretical
assumptions made to treat binding and Fermi motion corrections.

To summarise, the vertex functions that we use are given by
\begin{mathletters}
\label{verf}
\begin{eqnarray}
\Phi_1^S(p,k)
&\propto& { (k^2 - m^2) \over (k^2 - \Lambda_S^2)^2 }
          { (M^2 - \Lambda_p^2) \over (p^2 - \Lambda_p^2) }
                                                      \label{verfS}\\
\Phi_1^V(p,k)
&\propto& { (k^2 - m^2) \over (k^2 - \Lambda_V^2)^{7/2} }
          { (M^2 - \Lambda_p^2) \over (p^2 - \Lambda_p^2) }.
                                                      \label{verfV}
\end{eqnarray}
\end{mathletters}
We find the best fit to the experimental nucleon distributions at
$Q^2 = 4$ GeV$^2$ (we evolve the curves from $Q_0^2 = 0.15$ GeV$^2$
using leading order QCD evolution, with $\Lambda_{QCD} = 250$ MeV
\cite{GRV}) for masses $m_S = 850$ MeV and $m_V = 1050$ MeV,
and cut-offs $\Lambda_S = 1.2$ GeV and $\Lambda_V = 1.0$ GeV,
which we fit to the recent parameterisations by Morfin and Tung,
and Owens \cite{DATAFIT}.
The fits to the $u_V + d_V$ valence quark distribution as well as the
valence $d_V/u_V$ ratio are shown in Figs.\ref{4} and \ref{5}
respectively.
It is remarkable that such simple forms for the vertex functions
reproduce the data so well.

Having parameterised the free nucleon vertices, we are now ready to
consider the specific cases of DIS from
the deuteron, from nuclear matter, and from dressed nucleons.
Throughout, we consider the isoscalar valence structure function,
$x W_T \propto x (u_V + d_V) = 3 x (q_0 + q_1)/2$, where $q_0$ and
$q_1$ are the quark distributions arising in connection with the
scalar and vector diquarks, respectively,
normalised so that their first moments are unity
(from the spin-flavour wavefunction of the proton we have
$d_V = q_1$ and $u_V = (q_1 + 3 q_0)/2$).

\section{Calculation of Composite Target Structure Functions}

\subsection{DIS From the Deuteron}

We examine nuclear DIS from a deuteron for several reasons.
Firstly, it is critical to know the size of the off-mass-shell
corrections to the deuteron structure function if ultimately the
nuclear EMC data (which usually measures the ratio of nuclear to
deuterium structure functions) is to be used to draw conclusions
about the differences between quark distributions in free nucleons
and those bound in nuclei.
Secondly, in the absence of high-statistics neutrino data, the
neutron structure function is often inferred from the deuteron
structure function using the naive assumption of additivity
of the bound proton and neutron structure functions.
Apart from the off-mass-shell effects which we consider here,
several other effects spoil this simple assumption.
For example, nuclear shadowing is important
as $x \rightarrow 0$ \cite{SHADOW}, and of course the
deuteron structure function extends beyond
$x_N = 1\ (x_N \equiv (M_D/M)\ x)$ to $x_N = M_D/M$.
Hence deviations from additivity occur over much of the range of $x$.
For a reliable extraction of the neutron structure function a
systematic computation of these effects is clearly necessary.

The calculation of DIS from the deuteron is more straightforward
and reliable than for heavier nuclei, since the relativistic
deuteron--nucleon vertex is reasonably well understood.
The treatment of the deuteron recoil state is simplified
by the fact that most of the time this will be an on-shell nucleon,
as this can be expected to dominate the contributions from processes
with a recoil $\Delta$ or Roper resonance, or a higher mass state.

The structure of the general $DNN$ vertex, with one nucleon on-shell,
was first derived by Blankenbecler and Cook \cite{BLANK60},
$\langle N | \psi_N | D \rangle
\propto
(\not\!p - M)^{-1}
\Gamma^D_{\alpha}\ \epsilon^{\alpha}\ {\cal C} \bar{u}^T({\cal P}-p)
$,
where the $DNN$ vertex function is \cite{ARNOLD80}
\begin{eqnarray}
\Gamma^D_{\alpha}(p^2)
&=& \gamma_{\alpha}\ F(p)\
 +\ \left( {1 \over 2} {\cal P}_{\alpha} - p_{\alpha} \right)\ G(p)\
 +\ {\not\!p - M \over M}
    \left[ \gamma_{\alpha}\ H(p)\
       +\ \left( {1 \over 2} {\cal P}_{\alpha} - p_{\alpha} \right)\
          {J(p) \over M}
    \right],
\label{DNN}
\end{eqnarray}
and ${\cal C}$ is the charge conjugation operator.
The functions $F, G, H$ and $J$ are related to the
$^3S_1, ^3D_1, ^1P_1$ and
$^3P_1$ deuteron wavefunctions, $u, w, v_s$ and $v_t$, respectively,
by
\begin{mathletters}
\label{FGHI}
\begin{eqnarray}
F(p) &=& \pi \sqrt{2 M_D}\ (2 E_p - M_D)
     \left( u({\bf p})
          - {w({\bf p}) \over \sqrt{2}}
          + \sqrt{ {3 \over 2} } {M \over |{\bf p}|} v_t({\bf p})
     \right)                                                     \\
G(p) &=& \pi \sqrt{2 M_D}\ (2 E_p - M_D)
     \left( {M \over E_p + M} u({\bf p})
          + {M\ (2 E_p + M) \over {\bf p}^2} {w({\bf p})
                                              \over \sqrt{2}}
          + \sqrt{ {3 \over 2} } {M \over |{\bf p}|} v_t({\bf p})
         \right)                                                \\
H(p) &=& \pi \sqrt{2 M_D}\ {E_p M \over |{\bf p}|}\
         \sqrt{ {3 \over 2} }\ v_t({\bf p})                   \\
J(p) &=& -\pi \sqrt{2 M_D}\ {M^2 \over M_D}
         \left( {2 E_p - M_D \over E_p + M} u({\bf p})
              - {(2 E_p - M_D) (E_p + 2 M) \over {\bf p}^2}
                {w({\bf p}) \over \sqrt{2}}
              + {\sqrt{3} M_D \over |{\bf p}|}\ v_s({\bf p})
         \right)
\end{eqnarray}
\end{mathletters}
where $E_p = \sqrt{M^2 + {\bf p}^2}$ and ${\bf p}$ is the off-shell
nucleon's three-momentum.
For the deuteron wavefunctions we use the model of Buck and Gross
\cite{BUCK79}, with a pseudo-vector $\pi$-exchange interaction.

For the spin-averaged deuteron hadronic tensor we therefore
need to evaluate the trace
\begin{eqnarray}
\sum_{\lambda} \epsilon^{*\alpha}(\lambda,{\cal P})\
               \epsilon^{\beta}(\lambda,{\cal P})\
{\rm Tr} \left[ (\not\!\!{\cal P}^T - \not\!p^T  + M)\
           {\cal C} \overline{\Gamma}^D_{\beta}(p^2)\
           (\not\!p + M)\ \chi_{\mu\nu}(p,q)\ (\not\!p + M)\
           {\cal C} \Gamma^D_{\alpha}(p^2)
    \right]                                     \label{Dtrace}
\end{eqnarray}
where $\epsilon^{\alpha}(\lambda,{\cal P})$ is the polarisation
vector for a deuteron with helicity $\lambda$,
and $\overline{\Gamma}^D_{\beta}
= \gamma_0\ \Gamma^{D \dagger}_{\beta}\ \gamma_0$.
This yields the following deuteron--nucleon functions:
\begin{mathletters}
\label{Adeut}
\begin{eqnarray}
A_0^D(p^2)
&=& M \left\{
4\ F^2
\left[ 4\ M^2\ +\ 2\ M_D^2\
       -\ \left(p^2-M^2\right)
     \left( - 2 + {p^2-M^2 \over M_D^2} \right)
\right] \right.                                        \nonumber\\
&-& 8\ F\ G
\left[ 4\ M^2\ -\ M_D^2\
 +\ {\left(p^2-M^2\right) \over 4 M^2}
      \left( 10\ M^2 - M_D^2 + 2\ p^2
          + {3\ M^4 - 2\ M^2 p^2 - p^4 \over M_D^2}
      \right)
\right]                                                \nonumber\\
&+& {G^2 \over M^2}
\left[ \left( 4\ M^2 - M_D^2 \right)^2
     - \left( p^2-M^2 \right)
       \left( 4\ M_D^2 - 5\ p^2 - 11\ M^2 + {2 p^4-2 M^4 \over M_D^2}
       \right)
\right]                                                \nonumber\\
&-& {\left( p^2-M^2 \right) \over M^2}
\left[ - 12\ H^2\ \left( p^2-M^2 \right)
       + 4\ F\ H\ \left( -5 M^2 -2 M_D^2 + p^2
                         + {\left( p^2-M^2 \right)^2 \over M_D^2}
                  \right)
\right.                                                \nonumber\\
& & \hspace*{2cm}
+\ \left( p^2 - {({\cal P} \cdot p)^2 \over M_D^2} \right)
   \left( {\left( p^2-M^2 \right) \over M^2}
          \left(- 4\ J^2 + 8\ H\ J \right)
+\ 16\ F\ J
\right.                                                 \nonumber\\
& & \hspace*{3cm}
\left.
\left.
\left.  +\ 16\ G\ H\
        +\ 8\ G\ J\ {\left( {\cal P} \cdot p - M^2 - p^2 \right)
                           \over M^2}
\right)
\right]
\right\}                                                    \\
A^D_{1\alpha}(p^2)
&=&
4\ F^2
\left[ \left( 4\ M^2\ +\ 2\ M_D^2
       \right)\ p_{\alpha}
    -\ \left( p^2-M^2 \right)
       \left( { (p^2-M^2) \over M_D^2 } p_{\alpha}
  + \left( 2 - {(p^2-M^2) \over M_D^2} \right) {\cal P}_{\alpha}
       \right)
\right]                                              \nonumber\\
&-& 8\ F\ G
\left[ \left( 4\ M^2\ -\ M_D^2
       \right)\ p_{\alpha}
    +\ (p^2 - M^2)
       \left( \left( 1 - {(p^2-M^2) \over M_D^2} \right) p_{\alpha}
            + {{\cal P} \cdot p \over M_D^2} {\cal P}_{\alpha}
       \right)
\right]                                              \nonumber\\
&+& {G^2 \over M^2}
\left[ \left( 4\ M^2 - M_D^2 \right)^2\ p_{\alpha}
     - \left( p^2-M^2 \right)
       \left( (M_D^2 - 4\ M^2)
              \left( 2 - {p^2-M^2 \over M_D^2} \right) p_{\alpha}
       \right.
\right.                                               \nonumber\\
& & \hspace*{7cm}
\left.
       \left. - 4 \left( p^2 - {({\cal P} \cdot p)^2 \over M_D^2}
                  \right) {\cal P}_{\alpha}
       \right)
\right]                                              \nonumber\\
&-&
{ \left( p^2-M^2 \right) \over M^2 }
\left\{
  4\ H^2\ (p^2-M^2)
\left[ p_{\alpha} - \left( 2 - {p^2-M^2 \over M_D^2} \right)
                                            {\cal P}_{\alpha}
\right]
\right.                                               \nonumber\\
& &
-\ 4\ J^2\ {\left( p^2-M^2 \right) \over M^2}
\left( p^2 - {({\cal P} \cdot p)^2 \over M_D^2} \right)
\left[ p_{\alpha} - {\cal P}_{\alpha} \right]\
+\ 8\ H\ J\ (p^2-M^2)
\left[ p_{\alpha} - {{\cal P} \cdot p \over M_D^2} {\cal P}_{\alpha}
\right]                                               \nonumber\\
& &
-\ 8\ F\ H\ M^2
\left[ 2\ p_{\alpha} + \left( 2 - {p^2-M^2 \over M_D^2} \right)
                                              {\cal P}_{\alpha}
\right]                                               \nonumber\\
& &
+\ 4\ F\ J
\left[ \left( 3 M^2 + p^2 - {(p^2-M^2)^2 \over M_D^2} \right)
                                                 p_{\alpha}
     - \left( p^2 + M^2 - {(p^2-M^2)^2 \over M_D^2} \right)
                                                {\cal P}_{\alpha}
\right]                                             \nonumber\\
& &
+\ 8\ G\ H
\left[ \left( M^2 + p^2 - {\cal P} \cdot p \right) p_{\alpha}
     + \left( p^2 - {p^2+M^2 \over M_D^2} {\cal P} \cdot p\ \right)
                                                    {\cal P}_{\alpha}
\right]                                                \nonumber\\
& &  \left.
+\ 8\ G\ J \left( p^2 -{({\cal P} \cdot p)^2 \over M_D^2} \right)
\left[ -2 p_{\alpha} + {\cal P}_{\alpha}
\right]
\right\}.
\end{eqnarray}
\end{mathletters}

As mentioned in the previous section, we constrain the cut-off
parameter $\Lambda_p$ in the quark--(off-shell) nucleon vertex
functions defined in Eqs.(\ref{verf}) by fitting our full,
$p^2$-dependent calculated distribution
to the experimental deuteron structure function, using the
lepton--deuteron data from NMC, BCDMS and SLAC \cite{NMC92}.
However, we still need to fix the normalisation constants in
Eqs.(\ref{verf}).
Naturally, these will be functions of the cut-off
$\Lambda_p$.
Actually, if the exact quark--nucleon vertex functions were known,
they would be the same for the off-shell as for the on-shell nucleon.
We do not assume this, however, as the vertex functions which we use
are only approximations to the exact results.
For example, the arguments given in Section V, relating to the
counting rules which give the $k^2$ dependence of the vertex
functions, are based
on quark distributions in an on-shell nucleon.
In an off-shell nucleon the connection between $x$ and $k^2$ is
given by the modified expression
\begin{eqnarray}
k^2 &=& k_+ k_- - k_T^2\
     =\ x M_D \left( p_- - { ({\bf p}_T - {\bf k}_T)^2 + m_R^2
                             \over M_D (y-x) }
              \right)\
     -\ k_T^2,                                \label{k2off}
\end{eqnarray}
with $p_-$ now constrained by the $\delta$-function for the
on-shell nuclear recoil state (see Section IV).
In principle, the asymptotic $k^2$ dependence for the
quark--off-shell nucleon vertices expected from counting
rules could be determined after integration over
the nucleon's momentum.
Clearly this is a much more complicated task than was the case for
the on-shell nucleon, and we do not believe our simple ansatz for
the vertices warrants such a treatment, in which case we shall
simply normalise by comparing with the data.

In Fig.\ref{6} we compare the experimental
$F_{2D}$ at $Q^2 = 10$ GeV$^2$ with the
calculated total valence quark distribution in the deuteron,
$5 x(u_V+d_V)/9$, evolved from the same value of $Q_0^2 = 0.15$
GeV$^2$
(since we use the same diquark masses) as for the free nucleon
distributions in Section V.
The result of the full calculation is almost independent of the
value of $\Lambda_p$ used, after the normalisation constants for the
vertex functions have been determined by the charge conservation
condition.
This is because the $p_T$ distribution is strongly peaked at small
transverse momenta, $p_T \sim 25$ MeV, so that modification of the
large $p_T$ (or large $|p^2|$) behaviour by altering the form factor
cut-off has negligible consequences.
Clearly there is very good agreement between the model calculation
and the data for $x \agt 0.3$.

{}From the discussion in Sections IV and V it should be clear that
it is not possible to justify the convolution model for deuteron
deep-inelastic scattering.
Still, it is of interest to compare our results with those
of previous calculations that have made use of convolution-like
formulas.
Firstly we can notice that by taking the on-shell limit
($p^2 \rightarrow M^2$)
for the kinematic factors in $A_0^D$ and $A_1^D$
in Eqs.(\ref{Adeut}), we obtain
$A_0^D / M = p\cdot A_1^D / M^2 = q\cdot A_1^D / p\cdot q$,
thereby satisfying condition (c) in Section IV for the
convolution model (although this approximation need not be taken
in the functions $F, G, H, J$ themselves).
Such an approximation is in the spirit of that used in
Ref.\cite{GROSS92} for the nuclear structure functions.
The result of this approximation is shown in the dashed curve of
Fig.\ref{6}, where we have used the same normalisation constants in
Eqs.(\ref{verf}) (for $\Lambda_p = \infty$) as those determined
in the full calculation.
The effect is a reduction in the absolute value of the structure
function, without much affect on the shape.
By artificially normalising the new distribution so that the final
result conserves baryon number, this curve becomes almost
indistinguishable from the full result.
However, there is no good reason for using different
normalisation constants in this approximation, since the $p^2 = M^2$
limit is taken in the nuclear part of the diagram and thus should not
in principle affect the quark--off-shell nucleon vertex.

In other calculations using the convolution model for deuterium,
the most common prescription has been to drop all terms but
$I \chi_T^0$ in the expansion of $\chi_{\mu\nu}$
(in Eq.(\ref{Dtrace})),
and to replace $\chi_T^0$ by the experimental, on-shell structure
function of the nucleon \cite{KUSNOMOR,NAKWONG},
In Fig.\ref{6} the dotted curve shows the result after renormalisation
to ensure baryon number two for the deuteron.
It is somewhat surprising that the difference in shape
between the full result and this ansatz is as small as it is.
Still, a discrepancy of $\sim 20\%$ is quite significant in a system
as loosely bound as the deuteron.

A numerically significant difference between the convolution
approach and the exact calculation is of particular importance if
one recalls that the neutron structure function is extracted from
structure functions of light nuclei, such as deuterium,
using the convolution model.
Indeed, in view of the problems which we have just described,
it is rather worrying than our knowledge of $F_{2n}$ is based on
this.
As seen in Fig.\ref{6}, depending on the approximation or
ansatz taken in calculating $F_{2D}$, the deviation from the correct,
$p^2$-dependent result, will vary.
Still, although unsatisfactory from a theoretical point of view,
by artificially re-normalising the deuteron structure function
by hand so that it respects baryon number conservation, the
differences can be reduced.

A similar situation arises in calculations of the nuclear EMC effect,
in which differences between nuclear and deuteron structure functions
are explored.
Clearly for any accurate description of this effect we need firstly
to have a reliable method of calculating the deuteron structure
function.
As we have seen, the off-shell effects that are ignored in the
deuteron may be compensated for by suitably renormalising the final
result.
Whether this can also be done in other, heavier, nuclei is not clear.
Certainly in heavy nuclei we would expect off-shell effects
to play some role.
To date these have not been adequately accounted for,
and this is what we turn to next.

\subsection{Nuclear Matter}

For any nucleus we can easily repeat the above calculation
if we know the relativistic nucleon--nucleus vertex functions.
Unfortunately, at the present time these are not at all well known
for heavy nuclei.
A solution to this problem would be to simply parameterise the vertex
functions, and to make some assumptions for the nuclear recoil state.
Alternatively, if one tried to use non-relativistic nuclear models
as an approximation, it would be difficult to incorporate the
off-shell nucleon structure.
The best way is to consider first the simpler case of a nucleon
embedded in nuclear matter.
In this type of calculation the off-shell effects are parameterised
in the effective nucleon mass, $M \rightarrow M^*$.

Experimentally, the effective nucleon mass at nuclear matter
density ($\sim 0.15$ fm$^{-3}$) is found to be $\sim 0.7\ M$
\cite{QHD}.
Theoretically, there is a large number of models for nuclear matter,
which predict a wide range of effective nucleon masses.
The Quantum Hadrodynamics model of Walecka and Serot \cite{QHD},
in which
pointlike nucleons (in the mean field approximation) are bound by the
exchange of scalar ($\sigma$) and vector ($\omega$) mesons, predicts
rather small effective masses, $M^*/M \simeq\ 0.5-0.6$.
Somewhat larger masses are obtained when explicit quark degrees of
freedom are introduced.
For example, in the Guichon model \cite{GUICH}, where the $\sigma$
and $\omega$ mesons are allowed to couple directly to quarks inside
the nucleons,
the value of $M^*$ is typically $\sim 0.9\ M$.
Even larger values are obtained if one includes centre-of-mass
corrections
and self-coupling of the scalar fields \cite{SMT,FBSY90}.
Rather than choose a specific nuclear model, we let $M^*$ be a
parameter and examine the effect of its variation upon the nucleon
structure function,
defined in Eq.(\ref{WTreal}).

Because the quark--nucleon vertex function will now also depend
on the effective mass, it would be inappropriate to use the same
normalisation
constants in Eqs.(\ref{verf}) as those determined by normalising
the on-shell
nucleon distributions.
Therefore the normalisation constants in this case must be determined
by normalising the calculated quark distributions in nuclear matter,
for $p^2 = M^{*2}$, so that their first moments are unity.

Fig.\ref{7} shows the isoscalar valence nucleon structure function,
$x \left( u_V(x,p^2=M^{*2})\right.$
$\left.+ d_V(x,p^2=M^{*2}) \right)$,
for a range of effective masses, $M^*/M \sim 0.5 - 1$.
There is clearly quite significant softening of the structure
function, with the most prominent effects appearing for
$0.2 \alt x \alt 0.6$.

However, it should be remembered that our formalism neglects
interactions between the spectator quarks and the surrounding
nucleons in the nuclear
medium (i.e. it assumes the impulse approximation).
This has been found to be quite a poor approximation \cite{SMT} for
nuclear matter.
A simple way to estimate the importance of final state interactions
is to assume that the strength of the interaction of the spectator
diquark with the nuclear medium is 2/3 that of the nucleon
interaction,
and that it is independent of the mass of the diquark.
In that case the diquark mass is modified by $m_R \rightarrow m_R^*$,
where
\begin{eqnarray}
m_R^* &=& m_R - {2 \over 3} \left( M - M^* \right)
\end{eqnarray}
for both scalar and vector diquarks.
The effect of this is shown in Fig.\ref{8}.
As can be seen, interactions of the spectator diquark lead to a
hardening of the quark distribution, typically of the same order
of magnitude as the nucleon off-shell effects.
Combined with the off-shell effects, this gives a structure function
which (for $M^*/M \approx 0.7$) is $\sim 20-30\%$ larger than the
on-shell result for $x \agt 0.4$.
For quantitative comparison against deep-inelastic scattering data on
nuclear structure functions it would therefore be very valuable to
develop a consistent formalism incorporating both effects.

\subsection{DIS From Dressed Nucleons}

Models of the nucleon which incorporate PCAC by including a pion
cloud have been used in DIS, among other things, to estimate the size
of the $\pi NN$ form factor \cite{PINNFF}, and to calculate the
flavour symmetry breaking in the proton sea,
the possibility of which was recently suggested by the result of
the New Muon Collaboration's measurement of the Gottfried sum rule
\cite{NMC91}.
Previous covariant calculations \cite{PREVPION}
have all relied upon the same assumptions as for the nuclear
calculations,
namely the validity of the convolution formula in the first place,
and the lack of any dependence of the bound nucleon structure
function on $p^2$ and ${\bf p}_T$.
In this Section we apply the formalism we have developed for dealing
with off-shell effects to the part of this problem where the virtual
photon hits the virtual nucleon, with its spectator pion left
on-mass-shell.

In order to calculate this contribution the only additional
ingredient which we need is the `sideways' $\pi N N$ form factor,
$\Gamma_{\pi NN}(p^2)$,
where one nucleon is off-mass-shell. For this we use the same
monopole form that is usually used in the literature \cite{MULDERS92}
(see also \cite{SIDEW})
\begin{eqnarray}
\Gamma_{\pi NN}(p^2)
&\propto& \left( p^2 - \Lambda^2_{\pi N} \right)^{-1},
\end{eqnarray}
with $\Lambda_{\pi N} \sim 1.4$ GeV and a pseudoscalar
$\pi N$ coupling.
With this, we can rearrange the relevant trace in Eq.(\ref{ccc}),
\begin{eqnarray}
{\rm Tr} \left[ (\not\!\!{\cal P} + M)\ i \gamma_5
                \Gamma_{\pi NN}(p^2)\
               (\not\!p + M)\ \chi_{\mu\nu}(p,q)\ (\not\!p + M)\
               i \gamma_5 \Gamma_{\pi NN}(p^2)
        \right],                                      \label{pitrace}
\end{eqnarray}
to obtain the nucleon--pion functions $A_{0,1}^{\pi N}$,
\begin{mathletters}
\label{Api}
\begin{eqnarray}
A_0^{\pi N}(p,{\cal P}) &=&
\left( - m_{\pi}^2\ M \right) \Gamma^2_{\pi NN}(p^2)           \\
A^{\pi N}_{1\ \alpha}(p,{\cal P}) &=&
\left( -m_{\pi}^2\ p_{\alpha}\
       +\ (p^2 - M^2) (p_{\alpha} - {\cal P}_{\alpha})
\right) \Gamma^2_{\pi NN}(p^2)
\end{eqnarray}
\end{mathletters}
Again, as was the case for the deuteron, by inserting
$p^2 \rightarrow M^2$ in
$A_{1 \alpha}^{\pi N}$ we can satisfy the conditions of case (c)
in Section IV.
However, the structure function this time is proportional to
$- m_{\pi}^2$
(i.e. negative), which is clearly unphysical. This illustrates the
fact that even though the one-dimensional convolution formula may
indeed be obtained from the exact result by certain approximations
(e.g. on-shell limit),  there
is no guarantee that these approximations are physically meaningful.

As was shown in \cite{MULDERS92}, the convolution model may be
derived if, amongst other things, one assumes that the off-shell
nucleon structure is the same as that of a point-like fermion
\cite{JAFFE85}, in which case the relevant
operator in $\chi_{\mu\nu}$ is $\not\!q \chi^2_T$.
As we have seen above,
this is only part of the complete expression in the Bjorken limit
if one assumes the nucleon quark vertex to be of the form
in Section V.
Nevertheless, the model of \cite{MULDERS92} can be obtained
using these vertices if the
following steps are taken: firstly the trace in Eq.(\ref{pitrace})
evaluated
with the $\not\!q \chi_T^2$ structure; then to obtain factorisation
the limits
${\bf p}_T = 0$ and $p^2 = M^2$ taken in the
`nucleon structure function'
(i.e. $k$-dependent) parts; and finally the full structure of
the on-shell nucleon structure function used, as in Eq.(\ref{WTreal}),
rather than just keeping the $\chi_T^2$ term.
The necessity of the last point is clear, since for the on-shell
structure function the {\em individual} functions $\chi_T^i$ are not
necessarily positive definite, although the sum of course is
positive.

Other authors \cite{TEGEN} have implicitly assumed that
the relevant operator
to be used in the $\chi_{\mu\nu}$ of Eq.(\ref{pitrace})
is $I \chi_T^0$,
similar to what was done in the convolution model calculation
for the deuteron discussed in Section VI A.
However, even with the subsequent replacement of
$\chi_T^0$ by the full on-shell nucleon structure function in
the convolution expression, the result will be proportional
to $- m_{\pi}^2$ since the
coefficient of $\chi_T^0$ is $A_0^{\pi N}$.
Thus it appears that the result of
\cite{TEGEN} can only be obtained by taking the modulus of
a negative structure function.

Clearly, the above procedures are somewhat arbitrary.
It is a reflection of the fact that none of the scenarios
described in Section IV (namely cases (a)---(c)) for obtaining
the convolution model are applicable.
As in the deuteron case, the convolution model for dressed nucleons
is not derivable from the exact result.

In Fig.\ref{9} we show the result of the convolution model of
\cite{MULDERS92}.
This is compared with the result of the calculation including
the full $p^2$ dependence, with the quark--nucleon vertex
function evaluated with
$\Lambda_p = \infty$, as for the deuteron.
For the full calculation we use the same normalisation constants
for the quark--nucleon vertices as determined from the on-shell
nucleon calculation in Section V.
The results indicate that the full, $p^2$-dependent calculation
gives somewhat smaller results compared with those of the
convolution model
(although the shapes are quite similar, as can be seen from the
dotted curve, where we normalise the scalar and vector vertex
functions to give the same first moments
as in the convolution model).
Such a difference might have been surprising had the convolution
expression been a simple approximation to the full result,
in which case we may well have expected small off-shell corrections.
Unfortunately, this calculation is more difficult to check since
there is no clear normalisation condition for the structure
function.
Comparing the first moment of the calculated distributions with
the average number of pions in the intermediate state, which can be
calculated by considering DIS from the virtual pion, is ambiguous
due to the presence of antiparticles in the covariant formulation.
(A convolution formula such as Eq.(\ref{con1}) can be written for
DIS from virtual pions, since there are no spinor degrees of freedom
to spoil this factorisation. However, ambiguities in the $p^2$
dependence of the `off-shell pion structure function' would still
remain.)
We therefore believe that this fact illustrates the absence of a
firm foundation for the covariant convolution model for DIS from
dressed nucleons (see \cite{TOPTIMF} for an alternative approach
to this calculation).

\section{Conclusion}

We have investigated within a covariant framework the deep-inelastic
scattering from composite particles containing virtual nucleon
constituents.
The scattering has been treated as a two-step process, in which
the off-shell
nucleon in the target interacts with the high energy probe.
The treatment amounts to neglecting final state interactions.
We have constructed the truncated photon--nucleon amplitude from
14 general, independent functions, and used the parton model
to show that only 3 of these are relevant in describing the
deep-inelastic structure functions in the Bjorken limit.
The calculation explicitly
ensures current conservation and the Callan-Gross relation.

Within this framework we can unambiguously examine under what
conditions the conventional convolution model breaks down.
Furthermore, we use some simple models
of the relativistic quark--nucleon and nucleon--nucleus vertex
functions to investigate this breakdown numerically.
While the failure of the convolution model may
appear to be an unwelcome complication,
it is clear that in any theoretically self-consistent calculation
which takes off-mass-shell effects into account it is an inevitable
one.
Indeed, the `bound nucleon structure function' is an ill-defined
quantity within a covariant formulation.
This has wide-ranging consequences, as almost all calculations of
composite target structure functions (e.g. nuclei, for the EMC
effect) have
relied upon the validity of the simple convolution model.

We have been able to calculate the deuteron structure function
without making any assumptions about the $p^2$ dependence of the
structure functions, and find excellent agreement with the data in
the region of $x$ where our model is applicable ($x \agt 0.3$).
Making various assumptions for the off-shell nucleons naturally
introduces deviations from the exact result.
However, by suitably renormalising the approximated curves by hand
to ensure baryon number conservation (as was done in most previous
calculations) the differences between the exact
results and those of the convolution ansatz are minimised.
Although this is most unsatisfactory from a theoretical
point of view, phenomenologically the consequences of neglecting the
nucleon off-shell effects in the deuteron may not be too great.

To understand the consequences of the off-shell effects in heavy
nuclei, we considered a simple model of a nucleon embedded in
nuclear matter.
We found quite a significant softening of the structure function
at intermediate $x$ when the nuclear medium acts to decrease the
effective nucleon mass.
However, interactions of the spectator diquark state with the
surrounding medium tend to make the overall structure function
some $20-30\%$ harder at large $x$ ($x \agt 0.4$), for
$M^*/M \approx 0.7$, compared with the
on-shell result.

The other application which has been examined is DIS from the virtual
nucleon component of a physical, or dressed, nucleon,
where we also find quite significant differences between the full
result and the convolution model.
A detailed quantitative understanding of this effect is
needed in order to be able to describe the $x$ distributions for all
processes where the nucleon's dissociation into a virtual nucleon
and meson is expected to be of importance, such as in the
measurement of the asymmetry
in the light sea quark sector of the proton and neutron,
as well as the neutron spin structure function $g_{1n}(x)$.

\acknowledgements

This work was supported by the Australian Research Council.
One of us (AWS) would like to acknowledge the kind hospitality
extended to him by the theory group during his visit to Adelaide,
where this work was commenced.

\appendix

\section*{Gauge invariance and the longitudinal structure function}

Here we give the full details regarding the vanishing of the
longitudinal and non gauge-invariant structure functions.

Using the projection operators defined in Section II we project from
the truncated nucleon tensor $\chi_{\mu\nu}$ the contributions to
the longitudinal ($W^A_L$) and non gauge-invariant
($W^A_Q, W^A_{QL}$)
functions:
\begin{mathletters}
\label{prwhL}
\begin{eqnarray}
P^{\mu\nu}_L({\cal P},q)\ \chi_{\mu\nu}(p,q)
&=& \chi^0_L(p,q)\
 +\ \not\!p\ \chi^1_L(p,q)\
 +\ \not\!q\ \chi^2_L(p,q)\
 -\ {2\ p \cdot q \over q^2} \not\!q\ \chi^3(p,q)\      \nonumber\\
&+&
{q^2 \over (p \cdot q)^2}\ \left( p - y {\cal P} \right)^2
\left[ \chi^0_T(p,q)\
 +\ \not\!p\ \chi^1_T(p,q)\ +\ \not\!q\ \chi^2_T(p,q)\
\right]                                                 \\
P^{\mu\nu}_Q({\cal P},q)\ \chi_{\mu\nu}(p,q)
&=& \chi^0_Q(p,q)\
 +\ \not\!p\ \chi^1_Q(p,q)\
 +\ \not\!q\ \chi^2_Q(p,q)\                             \nonumber\\
&+& {2\ p \cdot q \over q^2} \not\!q\ \chi^3(p,q)\
 +\ 2\ \not\!q\ \chi^4(p,q)                             \\
- \frac{1}{2} P^{\mu\nu}_{QL}({\cal P},q)\ \chi_{\mu\nu}(p,q)
&=& \chi^0_{QL}(p,q)\ +\ \not\!p\ \chi^1_{QL}(p,q)\
    +\ \not\!q\ \chi^2_{QL}(p,q)\                       \nonumber\\
&+& {2\ p \cdot q \over q^2} \not\!q\ \chi^3(p,q)\
    +\ \not\!q\ \chi^4(p,q).
\end{eqnarray}
\end{mathletters}
Following the procedure described in Section III we find that
all the $\chi$'s in Eqs.(\ref{prwhL}) are of order $1/\nu$.
Hence in the Bjorken limit Eqs.(\ref{prwhL}) become
\begin{mathletters}
\label{prwhsL}
\begin{eqnarray}
P^{\mu\nu}_L({\cal P},q) \chi_{\mu\nu}(p,q)
&=& \not\!q\ \left( \chi^2_L(p,q)\
   - {2\ p \cdot q \over q^2} \chi^3(p,q) \right)       \\
P^{\mu\nu}_{Q}({\cal P},q)\ \chi_{\mu\nu}(p,q)
&=& \not\!q\ \left( \chi^2_Q(p,q)\
 +\ {2\ p \cdot q \over q^2} \chi^3(p,q)\
 +\ 2\ \chi^4(p,q) \right)                              \\
- \frac{1}{2} P^{\mu\nu}_{QL}({\cal P},q)\ \chi_{\mu\nu}(p,q)
&=& \not\!q\ \left( \chi^2_{QL}(p,q)\
 +\ {2\ p \cdot q \over q^2} \chi^3(p,q)\
 +\ \chi^4(p,q) \right).
\end{eqnarray}
\end{mathletters}
Furthermore, for the functions $\chi(p,q)$ we find that at
leading order in $\nu$,
\begin{mathletters}
\label{rel}
\begin{eqnarray}
\chi^2_L(p,q)
&=& {2\ p \cdot q \over q^2} \chi^3(p,q)        \\
\chi^2_Q(p,q)
&=& -{2\ p \cdot q \over q^2} \chi^3(p,q)\
    -\ 2\ \chi^4(p,q)                           \\
\chi^2_{QL}(p,q)
&=& \chi^2_Q(p,q) + \chi^4(p,q)                 \\
\chi_{L,Q,QL}^{0,1}(p,q)
&=& 0.
\end{eqnarray}
\end{mathletters}
Substituting these expressions into Eqs.(\ref{prwhsL}) therefore
leads to vanishing results for each of the longitudinal and non
gauge-invariant functions.
This result is true independent of the production mechanism of
the off-shell particle, that is, independent of the functions
$A_{0,1}$ as defined in Section IV.
For the special case of an on-shell nucleon the longitudinal
and gauge non-invariant structure functions are
\begin{mathletters}
\label{Wreal}
\begin{eqnarray}
{M \over 2} W^N_L(p,q)
\ =\ M\ \widetilde{\chi}^0_L(p,q)
&+& M^2\ \widetilde{\chi}^1_L(p,q)\
 +\ p \cdot q\ \widetilde{\chi}^2_L(p,q)\           \nonumber\\
&-& {2\ (p \cdot q)^2 \over q^2}\ \widetilde{\chi}^3(p,q)\
\rightarrow\  0                                           \\
{M \over 2} W^N_Q(p,q)
\ =\ M\ \widetilde{\chi}^0_Q(p,q)\
&+& M^2\ \widetilde{\chi}^1_Q(p,q)\
 +\ p \cdot q\ \widetilde{\chi}^2_Q(p,q)\           \nonumber\\
&+& {2\ (p \cdot q)^2 \over q^2}\ \widetilde{\chi}^3(p,q)\
 +\ 2\ p \cdot q\ \widetilde{\chi}^4(p,q)\
\rightarrow\ 0                                           \\
{M \over 2} W^N_{QL}(p,q)
\ =\ M\ \widetilde{\chi}^0_{QL}(p,q)\
&+& M^2\ \widetilde{\chi}^1_{QL}(p,q)\
 +\ p \cdot q\ \widetilde{\chi}^2_{QL}(p,q)\        \nonumber\\
&+& {2\ (p \cdot q)^2 \over q^2} \widetilde{\chi}^3(p,q)\
 +\ p \cdot q\ \widetilde{\chi}^4(p,q)\
\rightarrow\ 0
\end{eqnarray}
\end{mathletters}
where the zero results follow directly from (\ref{rel}).


\begin{figure}
\caption{The truncated nucleon tensor $\chi_{\mu\nu}$.}
\label{1}
\end{figure}

\begin{figure}
\caption{Scattering from an off-shell nucleon in a composite target.
        The functions $A_i$ describe the nucleon--composite target
        interaction.}
\label{2}
\end{figure}

\begin{figure}
\caption{Leading twist contribution to the off-shell tensor
         $\chi_{\mu\nu}$.
         The function $H(p,k)$ describes the soft, non-perturbative
         physics.}
\label{3}
\end{figure}

\begin{figure}
\caption{Valence $u_V+d_V$ quark distribution in the nucleon,
         evolved from $Q_0^2 = 0.15$ GeV$^2$
         (dashed curve) to $Q^2 = 4$ GeV$^2$ (solid curve),
         and compared against parameterisations (dotted curves)
         of world data \protect\cite{DATAFIT}.}
\label{4}
\end{figure}

\begin{figure}
\caption{Valence $d_V/u_V$ ratio,
         evolved from $Q_0^2 = 0.15$ GeV$^2$
         (dashed curve) to $Q^2 = 4$ GeV$^2$ (solid curve),
         and compared against parameterisations (dotted curves)
         of world data \protect\cite{DATAFIT}.}
\label{5}
\end{figure}

\begin{figure}
\caption{Valence part of the deuteron structure function:
         the solid line is the full calculation
         (with $\Lambda_p = \infty$);
         the dashed line is with the $p^2 = M^2$ approximation
         in $A_0, A_1$
         (case (c) in Section IV),
         with the same normalisation constants as in the full curve;
         the dotted line is the convolution model using only the
         $\chi^0_T(p,q)$ operator, together with the full nucleon
         structure function, normalised to baryon number one.
         The curves have been evolved from $Q_0^2 = 0.15$ GeV$^2$
         to $Q^2 = 10$ GeV$^2$ for comparison against the
         experimental
         $F_{2D}(x,Q^2=10 $GeV$^2$) \protect\cite{NMC92}.}
\label{6}
\end{figure}

\begin{figure}
\caption{Nucleon structure function in nuclear matter, in the impulse
         approximation, for a range of effective nucleon masses,
         evolved from $Q_0^2 = 0.15$ GeV$^2$ to $Q^2 = 4$ GeV$^2$.}
\label{7}
\end{figure}

\begin{figure}
\caption{As in Fig.\protect\ref{7}, but including the effects of
         interaction of the spectator diquark with
         the nuclear medium.}
\label{8}
\end{figure}

\begin{figure}
\caption{Contribution to the structure function of a nucleon
         from DIS off a virtual nucleon with a pion in the
         final state.
         The convolution model of Ref.\protect\cite{MULDERS92}
         (dashed) is compared with the full calculation
         (for $\Lambda_p = \infty$),
         using the same normalisation for the quark--off-shell
         nucleon vertices as for the on-shell vertices (solid),
         and normalising the $p^2$-dependent scalar and vector
         vertex functions (dotted) to give the same
         first moments as in the convolution model.
         All curves are evolved from $Q_0^2 = 0.15$ GeV$^2$
         to $Q^2 = 4$ GeV$^2$.}
\label{9}
\end{figure}

\end{document}